\documentclass[aps,pra,twocolumn,longbibliography,footinbib,groupedaddress]{revtex4-1}
\usepackage{graphicx}
\usepackage{multirow}
\usepackage{amsmath}
\usepackage{amssymb}
\usepackage[utf8]{inputenc}
\usepackage{array}
\usepackage{color}
\usepackage[caption=false]{subfig}

\begin{document}

\title{Mollow triplet in cold atoms}

\author{Luis Ortiz-Guti\'errez$^{1,2}$}
\author{Raul Celistrino Teixeira$^{3}$}
\author{Aur\'elien Eloy$^{1}$}
\author{Dilleys Ferreira da Silva$^{1,2}$}
\author{Robin Kaiser$^{1}$}
\author{Romain Bachelard$^{1,3}$}
\author{Mathilde Fouch\'e$^{1}$}    \email{mathilde.fouche@inphyni.cnrs.fr}

\affiliation{$^{1}$Universit\'e C\^ote d'Azur, CNRS, INPHYNI, France}
\affiliation{$^{2}$CAPES Foundation, Ministry of Education of Brazil, Caixa Postal 250, Brasília – DF 70040-020, Brazil}
\affiliation{$^{3}$Departamento de F\'{\i}sica, Universidade Federal de S\~{a}o Carlos, Rodovia Washington Lu\'{\i}s, km 235 - SP-310, 13565-905 S\~{a}o Carlos, SP, Brazil}

\date{\today}

\begin{abstract}
{In this paper, we measure the spectrum of light scattered by a cold atomic cloud driven by a strong laser beam. The experimental technique is based on heterodyne spectroscopy coupled to single-photon detectors and intensity correlations. At resonance, we observe the Mollow triplet. This spectrum is quantitatively compared to the theoretical one, emphasizing the influence of the temperature of the cloud and the finite-size of the laser beam. Off resonance measurements are also done showing a very good agreement with theory.}

\end{abstract}


\maketitle


\section{Introduction} \label{Sec:Intro}

The Mollow triplet, described theoretically by B. R. Mollow in 1969\,\cite{Mollow_1969}, corresponds to the three peaks of inelastically scattered light in the fluorescence spectrum of a two-level system driven by a strong and resonant incident field. Besides its importance in describing the atom fluorescence, it can also be viewed as a fundamental signature of quantum optics, highlighted by photon correlations between the peaks of the spectrum and antibunching in some particular cases. These correlations were first observed by A. Aspect \textit{et al.} in 1980\,\cite{Aspect_1980} and quantitatively characterized by C. A. Schrama \textit{et al.} in 1992\,\cite{Schrama_1992} which had remained the state of the art until recently. Indeed, this topic has attracted a renewed interest\,\cite{Ulhaq_2012}, mainly in solid state physics where, for example, Mollow triplet in a quantum dot appears as a promising candidate of heralded single-photon sources\,\cite{Lopez_2017}.

This power spectrum was theoretically investigated in the 60's and it was first established by B. R. Mollow\,\cite{Mollow_1969}. It can be decomposed in an elastic component, represented by a delta function in frequency space, and an inelastic component. The latter, in the specific case of resonant and strong enough driving field, is commonly known as the Mollow triplet: the spectrum is made of a carrier and two spectrally symmetric sidebands. It can be interpreted as the signature of spontaneous emission down a ladder of paired states that corresponds to the new eigenstates of the two-level system dressed by a strong driving field\,\cite{CCT_1977}.

The Mollow triplet was first observed in atomic beams\,\cite{Schuda_1974,Wu_1975,Grove_1977,Hartig_1976,Schrama_1992}, with a configuration that allowed to avoid as much as possible the inhomogeneous broadening due to temperature. The spectrum was usually measured using a Fabry-Perot cavity. Several improvements have been made, especially regarding the polarization of the driving field, to be as close as possible to a two-level system. This led to a first quantitative agreement with theory in 1977\,\cite{Grove_1977}, despite the presence of unaccounted effects such as the inhomogeneous broadening due to the nonuniformity of the laser field. Since then, the Mollow triplet has been observed in many different systems: ions\,\cite{Stalgies_1996}, single molecules\,\cite{Wrigge_2007}, or quantum dots\,\cite{Flagg_2009,Ulhaq_2012,Peiris_2015,Lagoudakis_2017}.

Surprisingly, the Mollow triplet was rarely studied in cold atomic clouds despite its specific advantages. The low temperature makes its impact on the fluorescence spectrum minimal. Under certain conditions, atomic beam experiments gave unexpected asymmetric spectra\,\cite{Hartig_1976,Grove_1977}. This effect, due to position dependent atomic recoil\,\cite{Prentiss_1987}, is inherent to the specific configuration implemented to minimize the Doppler contribution of these atomic beams. Compared to single-particle systems, a multi-atom source increases the signal available for the detection. On the other hand, solid state quantum emitters often exhibit significant deviations from the ideal two-level system, for example through phonon coupling\,\cite{Roy_2011,Ulrich_2011}. And finally, cold atomic clouds can reach homogeneous optical densities of up to about 1000\,\cite{Hsiao_2018} which makes them suitable to go beyond the single scattering approximation and identify fundamental effects of multiple scattering in the Mollow triplet, including asymmetries and additional resonances\,\cite{Ott_2013,Pucci_2017}. One of the first attempts using cold atoms was done by K. Nakayama \textit{et al.}\,\cite{Nakayama_2010} on optical molasses. However, the Mollow spectrum was characterized out of resonance, where the spectrum is composed mainly of the Doppler-broadened elastic part plus two sidebands of inelastically scattered light. At resonance, the Mollow triplet was hardly visible and a quantitative comparison with the theoretical spectrum is still lacking. In 2016, K. M. Shafi \textit{et al.}\,\cite{Shafi_2016} also published some data that could be interpreted in terms of Mollow spectrum. However, the experiment was done on a magneto-optical trap (MOT), thus with an atomic cloud illuminated by six different off-resonant laser beams and in the presence of a magnetic field gradient, which makes the interpretation of the data and its quantitative analysis much more challenging.

In this paper, we present the measurement of the spectrum of light scattered by a cold atomic cloud and driven by a strong laser field. We detect this optical spectrum using a beat note (BN) technique, coupled with an intensity correlation measurement setup\,\cite{Eloy_2018}. This allows us to take advantage of the good sensitivity provided by the heterodyne spectroscopy and of photon counting devices suited for extremely low signal\,\cite{Hong_2006}. We report on both resonant and off-resonant excitation. In the first case, the Mollow triplet is clearly visible. We will show that the experimental data are in quantitative agreement with the expected spectra if one takes into account some inhomogeneous broadening. For off-resonance excitation, the amount of elastic and inelastic scattering is quantified and presents a very good agreement with theory. Finally, we investigate the effect of light polarization and intensity modulation to show the importance of being as close as possible to a two-level system with homogeneous excitation to observe a clear and reliable signal.


\section{Experimental setup} \label{sec:SetUp}

\subsection{Cold atomic cloud}

The experimental setup is described in Refs.\,\cite{Kashanian_2016,Eloy_2018}. A cold atomic cloud is provided by loading $^{85}$Rb atoms in a MOT. A compression stage is added to increase its density and to obtain a smooth density profile. For this specific experiment, the cloud is made of typically $5\times10^7$ atoms with a longitudinal rms radius $\sigma_\mathrm{z}=0.5\pm 0.1$\,mm, and a radial rms radius $\sigma_\mathrm{r}=0.7\pm0.1$\,mm after a free expansion of 2\,ms. The temperature measured by time of flight (TOF) technique is of the order of 100\,$\mu$K before applying the strong laser beams to detect the Mollow spectrum.

\subsection{Laser beams} \label{subsec:Laser}

\begin{figure}[h]
	\centering
	\includegraphics[width=8cm]{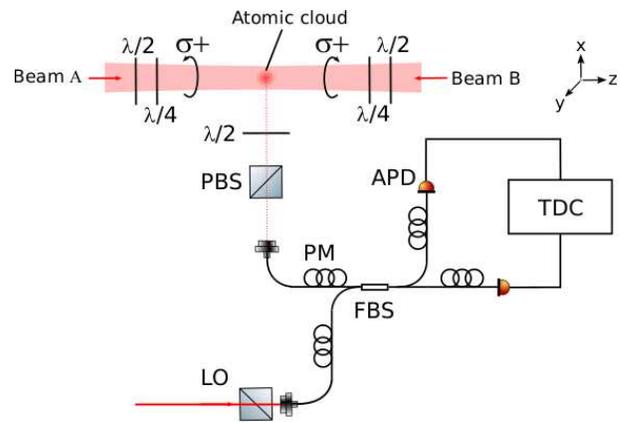}
	\caption{Experimental setup. Two counter-propagating laser beams (beams A and B) illuminate a cold atomic cloud. The spectrum of the scattered light is measured thanks to a beat note technique and its intensity autocorrelation. The scattered light is collected by a polarization maintaining (PM) single-mode fiber, after passing through a $\lambda/2$ waveplate and a polarizing beam splitter (PBS) to select the relevant polarization (see text). This fiber is connected to a fiber beam splitter (FBS) to mix the scattered light with the local oscillator (LO). The outputs of the FBS are sent to two avalanche photodiodes (APD). The single counts in each APD are time-tagged by a time-to-digital converter (TDC) and analyzed with a computer.}
	\label{fig:Setup}
\end{figure}

To observe the Mollow triplet, the atomic cloud is illuminated by two counter-propagating laser beams A and B, as depicted in Fig.\,\ref{fig:Setup}, applied along the longitudinal cloud axis $z$. Their power can be adjusted independently which allows, for instance, to avoid pushing the atoms due to imbalanced radiation pressure forces. The two beams come from the same distributed-feedback laser amplified by a tapered amplifier, and their frequency is locked close to the $F = 3 \rightarrow F' = 4$ hyperfine transition of the Rb$^{85}$ $D_2$ line, with linewidth $\Gamma=2\pi\times 6.07$\,MHz. The locking system uses a master-slave configuration with an offset locking scheme\,\cite{Puentes_2012}, allowing us to adjust the laser frequency without changing the beam direction and power.

We adjust the laser beam polarization, measured with a polarimeter just before the vacuum cell, with a $\lambda/2$ plate and a $\lambda/4$ plate. To probe a single closed transition, and thus consider the atoms as two-level systems,  we use the same circular polarization for the two beams. For a $\sigma^+$ polarization, the atoms are quickly pumped in ($F=3$, $m_F = 3$) and we probe the $F = 3\ (m_F = 3) \rightarrow F'=4\ (m'_F = 4)$ transition. Since both beams have the same polarization, if applied simultaneously, they would interfere and create a standing wave, leading to a strong spatial modulation of the intensity\,\cite{Fang_2016} with a corresponding convolution of the expected Mollow triplet. To avoid such a convolution, one need to apply a single laser beam with a well defined homogeneous Rabi frequency. We therefore illuminate the atoms alternating successively beam A and beam B, the switching on and off being done using two acousto-optic modulators (AOMs).

On the one hand, achieving a homogeneous Rabi frequency in the transverse direction would require a uniform intensity distribution, i.e., a plane-wave. On the other hand, the maximum available laser power imposes to decrease the beam waist to get a Rabi frequency as high as possible and thus well-resolved Mollow sidebands. As a compromise between the laser intensity and its non-uniformity over the cloud, we have used a beam waist of $w_0 = 2.5$\,mm on our cloud of radius 0.7\,mm. The laser beams are centered on the atoms using absorption imaging. We also observe an increase of the transverse radius of the cloud of about 10$\%$ after the alternating laser beams have been applied. The observed increase in the longitudinal direction is much higher, with typically an increase by a factor of 2.

Finally, the power of the laser beams can be adjusted to vary the Rabi frequency, by means of a $\lambda/2$ and a polarizing beam splitter (PBS). The maximum available power is about 20\,mW per beam, corresponding to a maximum on-resonance saturation parameter at the center of the laser beam of $s_{0,\mathrm{max}} = I/I_\mathrm{sat} \simeq  140$, with a saturation intensity $I_\mathrm{sat} = 1.67$\,mW/cm$^2$ for the transition considered in this experiment, and a maximum Rabi frequency of $\Omega_{\mathrm{max}} = \Gamma \sqrt{s_0/2} \simeq  8\,\Gamma$. We also use two photodiodes to monitor the power of each beam and their fluctuations during an experimental run which can last from a few hours to one day. We typically observe intensity fluctuations of 5\,$\%$.

\subsection{Measurement of the first order correlation function}

Our goal is to measure the spectrum $\tilde{g}_\mathrm{sc}{}^{(1)}(\omega)$ of the light scattered by the atomic cloud, or its Fourier transform which corresponds to the first order temporal correlation function $g_\mathrm{sc}{}^{(1)}(\tau)$. The latter quantity is obtained by recording the beat note between the scattered light and a local oscillator (LO). The light scattered while illuminating the atomic cloud with beam A or B is collected by a polarization-maintaining (PM) single-mode fiber. This allows us to select only one spatial mode. A $\lambda/2$ plate and a PBS placed in front of the collection fiber allows to select one polarization mode and to optimally couple the polarized scattered light into the fiber. Indeed, the angle between the laser beams and the axis of the fiber is close to 90$^\circ$. As the polarization of beams A and B are circular, once the atoms are pumped into the stretched $m_F=+3$ Zeeman sublevel, the polarization of the fluorescence in the direction of the fiber will be mostly linear. The $\lambda/2$ plate is thus adjusted to align the linear polarization of the scattered photons in order to maximise the transmission of the PBS. The local oscillator is derived from the same laser that provides beams A and B, but with an extra detuning $\omega_\mathrm{BN} = 2\pi \times 120$\,MHz. The local oscillator is then injected in another PM single-mode fiber, with a polarization selected by a PBS parallel to the fluorescence one.

The two fibers that collect the local oscillator and the light scattered by the atoms are then connected to a fiber beam splitter (FBS). We compute the $|g_\mathrm{sc}{}^{(1)}(\tau)|$ function via the measurement of the intensity autocorrelation of the beat note $g_\mathrm{BN}{}^{(2)}(\tau)$, given by the following formula\,\cite{Hong_2006}:
\begin{eqnarray}
g_\mathrm{BN}{}^{(2)}(\tau) &=& \frac{\langle I_\mathrm{BN}(t)I_\mathrm{BN}(t + \tau) \rangle}{\langle I_\mathrm{BN}(t) \rangle^2},\\
&=& 1 + 2\frac{\langle I_\mathrm{sc}(t) \rangle \langle I_\mathrm{LO}(t) \rangle}{\left(\langle I_\mathrm{sc}(t) \rangle + \langle I_\mathrm{LO}(t) \rangle \right)^2}|g_\mathrm{sc}{}^{(1)}(\tau)| \cos(\omega_\mathrm{BN} \tau) \nonumber\\
&& \frac{\langle I_\mathrm{sc}(t) \rangle^2}{\left(\langle I_\mathrm{sc}(t) \rangle + \langle I_\mathrm{LO}(t) \rangle \right)^2}\left(g_\mathrm{sc}{}^{(2)}(\tau)-1 \right), \label{eq:g2_BN}
\end{eqnarray}
with $I_\mathrm{BN}$, $I_\mathrm{sc}$, $I_\mathrm{LO}$ the intensity of the beat note, the collected scattered light and the local oscillator respectively, $g_\mathrm{sc}{}^{(2)}(\tau)$ the temporal intensity correlation of the scattered light, and with $\langle.\rangle$ corresponding to averaging over time $t$. It is important to note that this signal is not sensitive to the field correlation of the laser itself. This is due to the fact that the local oscillator and laser beams A and B are derived from the same laser, a well known technique used to get rid of the laser linewidth, at least at the first order. This is an advantage compared to other kind of techniques often used to measure the spectrum, such as Fabry-Perot interferometers, which require a laser with a low spectral width as well as a high resolution spectrometer\,\cite{Hartig_1976, Grove_1977, Flagg_2009}.

Experimentally, we have a local oscillator intensity more than 10 times higher than the scattered intensity. In this case, one can neglect the last term in Eq.\,(\ref{eq:g2_BN}). In addition, it is shifted in frequency by $\omega_\mathrm{BN}$ from the $g_\mathrm{sc}{}^{(1)}$ term under investigation. We are left with a signal oscillating at the frequency $\omega_\mathrm{BN}$ and with an amplitude proportional to $|g_\mathrm{sc}{}^{(1)}(\tau)|$. We then take the Fourier transform of $g_\mathrm{BN}{}^{(2)}(\tau)-1$ which contains the spectrum of the scattered light shifted by $\omega_\mathrm{BN}$.

The experimental setup to measure intensity correlations has been described in Ref.~\cite{Eloy_2018}. As shown in Fig.\,\ref{fig:Setup}, the two outputs of the FBS are sent to two single-photon avalanche photodiodes (APDs). This scheme allows to overcome the photodiode dead time and afterpulsing. The counts of the two APDs are time-tagged thanks to a time-to-digital converter (TDC). The corresponding file is finally sent to Matlab to calculate the histogram of the coincidences.

\subsection{Temporal experimental sequence}

To measure the Mollow triplet, the following time sequence is used. We first turn off the trapping laser beams and the magnetic field. The atoms are released from the MOT and a TOF of 2\,ms is applied to ensure that the MOT magnetic field gradient is completely off while keeping a small cloud radius compared to the waist of the incident laser beam. Then, we apply the laser beams. To avoid any interference effect, the counter-propagating beams A and B are switched on successively during $t_\mathrm{pulse} = 20\,\mu$s for each pulse, with a waiting time of 20\,$\mu$s between two consecutive pulses. We use for each run a train of 10 pulses, i.e. 5 pulses from beam A and 5 pulses from beam B. These parameters were chosen to minimize the effect of pushing and heating the atoms while keeping a reasonable pulse duration to perform a measurement.

The APDs are gated on the beam pulses. We calculate the histogram of the temporal coincidences for each pulse and after the first\,$\mu$s of the pulse once the laser intensity is stable. This also ensures that the stationary state is reached, as assumed in the calculations of the Mollow triplet\,\cite{Mollow_1969, Kimble_1976}. We then analyse the data coming from beams A and B separately. We first check that the histograms are the same for all the pulses, allowing to sum the 5 histograms corresponding to each beam. Because of the finite time-window of the measurement, the histogram needs to be normalized to get the intensity correlation function, as explained in Ref.\,\cite{Eloy_2018}.


\section{Results and Discussion} \label{Sec:Results}

\subsection{Theoretical spectrum} \label{Subsec:Theory}

The spectrum $\tilde{g}_\mathrm{sc}{}^{(1)}(\omega)$ of light scattered by a two-level system driven by a strong incident field close to resonance is commonly known as the Mollow triplet. Its derivation can be found in Ref.\,\cite{Mollow_1969} assuming a quantum two-level system driven by a classical field, or in Ref.\,\cite{Kimble_1976} with a fully quantum-mechanical approach. It can be decomposed in two parts:
\begin{equation}
\tilde{g}_\mathrm{sc}{}^{(1)}(\omega) = \tilde{g}_\mathrm{el}{}^{(1)}(\omega) + \tilde{g}_\mathrm{inel}{}^{(1)}(\omega), \label{eq:total_spectrum}
\end{equation}
with $\tilde{g}_\mathrm{el}{}^{(1)}(\omega)$ the elastic part and $\tilde{g}_\mathrm{inel}{}^{(1)}(\omega)$ the inelastic part given by the following formula:
\begin{widetext}
\begin{eqnarray}
&&\tilde{g}_\mathrm{el}{}^{(1)}(\omega) = \frac{s}{2(1+s)^2}\delta(\omega), \label{eq:el_spectrum}\\
&&\tilde{g}_\mathrm{inel}{}^{(1)}(\omega) = \frac{s_0}{8\pi\Gamma} \frac{s}{1+s}  \times \frac{ 1+ \frac{s_0}{4} +\left(\frac{\Delta}{\Gamma}\right)^2}{\left[\frac{1}{4}+\frac{s_0}{4} + \left(\frac{\Delta}{\Gamma}\right)^2 - 2\left(\frac{\omega}{\Gamma}\right)^2 \right]^2 + \left(\frac{\omega}{\Gamma}\right)^2\left[\frac{5}{4}+\frac{s_0}{2} + \left(\frac{\Delta}{\Gamma}\right)^2 - \left(\frac{\omega}{\Gamma}\right)^2 \right]^2}, \nonumber\\
&& \label{eq:Inel_spectrum}
\end{eqnarray}
\end{widetext}
where $\omega$ denotes the frequency relative to the atomic transition. The laser detuning is denoted by $\Delta$, and the saturation parameter depends on the laser detuning as follows:
\begin{equation}
s(\Delta) = \frac{\frac{I}{I_\mathrm{sat}}}{1+4\left(\frac{\Delta}{\Gamma} \right)^2} = \frac{s_0}{1+4\left(\frac{\Delta}{\Gamma} \right)^2}. \label{eq:s}
\end{equation}
One can show that the ratio of the intensity of the inelastic part to the elastic part, obtained by integrating Eqs.(\ref{eq:el_spectrum}-\ref{eq:Inel_spectrum}) over the emitted spectrum, is given by:
\begin{equation}
\frac{I_\mathrm{inel}}{I_\mathrm{el}} = \frac{s^2/(1+s)^2}{s/(1+s)^2} = s. \label{eq:I_inel_I_el}
\end{equation}
For $\Omega \gg \Gamma$ and $\Omega \gg \Delta$, the inelastic part can be approximated by three Lorentzians, the carrier at the laser frequency and the sidebands separated from the carrier by the generalized Rabi frequency:
\begin{equation}
\Omega_\mathrm{G} = \sqrt{\Omega^2 + \Delta^2}, \label{eq:Omega_Rabi_G}
\end{equation}
with $\Omega$ the Rabi frequency given by:
\begin{equation}
\Omega = \Gamma \sqrt{\frac{s}{2}\left[1+4\left(\frac{\Delta}{\Gamma} \right)^2 \right]} = \Gamma\sqrt{\frac{s_0}{2}}. \label{eq:Omega_Rabi}
\end{equation}
At resonance, the height ratio is 1:3:1 ans the width is $3\Gamma/2$ for the two sidebands and $\Gamma$ for the carrier, corresponding to an intensity (area) ratio of 1:2:1.

\subsection{First order correlation function and spectrum}

\begin{figure}[h]
	\centering
	\includegraphics[width=8cm]{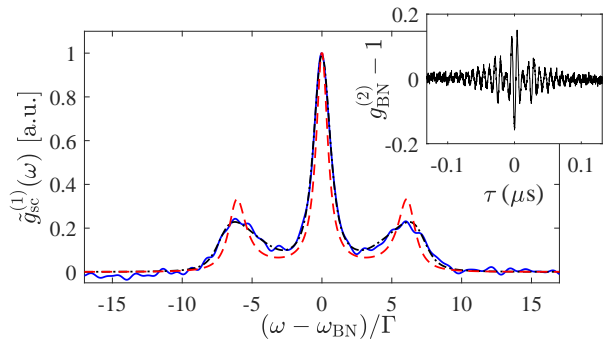}
	\caption{Resonant Mollow triplet of the light scattered by the cold atomic cloud, with no detuning and $s_0 \simeq 80$. Blue curve: experimental data; dashed black curve: fit by Eq.\,(\ref{eq:Inel_spectrum}); dash-dotted red curve: fit including the inhomogeneous broadening due to temperature and laser finite size. Inset: temporal intensity autocorrelation of the beat note.}
	\label{fig:Fig2_g1t_g1omega_b}
\end{figure}

A typical experimental intensity correlation is plotted in the inset of Fig.\,\ref{fig:Fig2_g1t_g1omega_b}. The laser frequency was set to resonance and with a saturation parameter of the order of 80. We first observe a fast oscillation at the frequency $\omega_\mathrm{BN}$ corresponding to the frequency beat note between the local oscillator and the light scattered by the atomic cloud. These oscillations decay on a time scale of the order of $1/\Gamma \simeq 26$\,ns. Finally, we can identify the beat note between different frequency components, namely between the Mollow sidebands and its carrier, giving rise to a revival of the amplitude of the oscillations.

In our experimental conditions, $g_\mathrm{BN}{}^{(2)}(\tau)-1$ is directly proportional to $|g_\mathrm{sc}{}^{(1)}(\tau)| \cos(\omega_\mathrm{BN}\tau)$. The Fourier transform, plotted in blue in Fig.\,\ref{fig:Fig2_g1t_g1omega_b}, will thus directly give the spectrum of the scattered light. The height of this spectrum has been arbitrarily normalized to unity. The Mollow triplet is clearly visible with its carrier and the two sidebands.

\subsection{Fitting procedure}

When $s_0 \gg 1$, the inelastic part is dominant and we thus neglect, at first, the elastic part. To extract $s_0$ from the experimental spectrum, and the corresponding Rabi frequency, we first fit the data using Eq.\,(\ref{eq:Inel_spectrum}). The result is plotted in dashed black line in Fig.\,\ref{fig:Fig2_g1t_g1omega_b}. The first observation is that the full width at half maximum (FWHM) of the carrier is underestimated: the experimental data yields a FWHM of $1.4\,\Gamma$ instead of the expected value $\Gamma$. We can also notice that the height ratio between the sidebands and the carrier is not 1:3:1 as expected but of the order of 0.7:3:0.7 and that the sideband widths are larger than expected.

This discrepancy can be explained by different contributions. We will first focus on the sidebands. The first contribution comes from the laser linewidth $\Delta\nu_\mathrm{L} = 3$\,MHz. As said before, this linewidth has no effect on the carrier because we use the same laser for the local oscillator and the laser beams A and B. On the other hand, different frequencies on the atoms lead to different generalized Rabi frequencies, and consequently to a broadening of the sidebands. However, if $\Delta\nu_\mathrm{L} \ll \Omega$, the change in the generalized Rabi frequency is of the order of $\Delta\nu_\mathrm{L}^2/2\Omega^2$. The minimum Rabi frequency used on this experiment is higher than 4\,$\Gamma$, corresponding to a change in\,$\Omega_\mathrm{G}$ of less than 1\,$\%$ which can thus be neglected. The same argument holds if one considers the temperature of the cloud $T$. The distribution of the effective detunings, in the atomic rest frame, corresponds to a Gaussian with an rms width of $k\sqrt{k_\mathrm{B}T/M}$, where $k$ is the wave vector, $k_\mathrm{B}$ is the Boltzmann constant and $M$ is the atomic mass. We will see in the next section that, when the laser beams A and B are applied, $T$ increases to a few tens of\,mK. This corresponds to a rms width of a few MHz, thus still with almost no effect on the generalized Rabi frequency and the sidebands broadening. The last contribution, which actually fully explains the sidebands broadening, comes from the finite waist of the laser beam. This results in a Rabi frequency not being the same for all the atoms. This does not affect the carrier width but leads to an inhomogeneous broadening of the sidebands: their height is thus decreased and their width increased.

Finally, the increase in the carrier width is due to temperature. In the previous paragraph, we have evaluated the impact of the frequency detuning of the incident photon in the atomic rest frame, but one also needs to take account the frequency  shift of the scattered photon. This effect has been characterized quantitatively in one of our previous experiments on the elastic spectrum\,\cite{Eloy_2018}, showing that the theoretical spectrum, given for the inelastic part by Eq.\,(\ref{eq:Inel_spectrum}) for atoms at rest, should be convoluted by the Doppler-broadened spectrum. This spectrum is Gaussian with a rms width $\Delta\omega_\mathrm{D}$ that depends on the temperature as well as the angle $\theta$ between the incident photon and the scattered direction:
\begin{equation}
\Delta\omega_\mathrm{D} = k \sqrt{2(1-\cos \theta)k_\mathrm{B}T/M}. \label{eq:Delta_omega_Doppler}
\end{equation}

To take into account these two contributions, namely temperature and finite waist of the laser beam, we have implemented a new fitting procedure. First, the laser beam is considered homogeneous in the propagation direction $z$ since the Rayleigh length is of several meters. Then, we model the cloud as a continuous Gaussian distribution which, after integration over the longitudinal axis $z$, reads $\rho(r_\perp)=N\exp(-r_\perp^2/2\sigma_r^2)/(2\pi\sigma_r^2)$. The spectrum radiated by the cloud in the finite-waist beam, with Rabi frequency $\Omega_{\mathbf{r}_\perp}=\Omega\exp(-r_\perp^2/w_0^2)$, is given by:
\begin{equation}
\tilde{g}_\mathrm{cloud}{}^{(1)}=2\pi\int_0^\infty \rho(\mathbf{r}_\perp) \tilde{g}_\mathrm{\Omega_{\mathbf{r}_\perp}}{}^{(1)}(\omega) r_\perp dr_\perp,\label{eq:GaussSpec}
\end{equation}
where the single-atom spectrum $\tilde{g}_\mathrm{\Omega_{\mathbf{r}_\perp}}{}^{(1)}$ is given by Eq.(\ref{eq:Inel_spectrum}) for a Rabi frequency $\Omega_{\mathbf{r}_\perp}$. Finally, to take into account the effect of the temperature, the spectrum is convoluted with the following Gaussian: $\exp(-\omega^2/2\Delta\omega_D^2)$ and the final height is renormalized to unity.
The free parameters are the frequency beat note $\omega_\mathrm{BN}$ (which is always in agreement with the experimental value, within fluctuations of $\sim1\%$), the Rabi frequency at the center of the beam $\Omega$, the ratio between the laser beam waist and the radial atomic cloud radius $w_0/\sigma_\mathrm{r}$, and the temperature $T$.

\subsection{Results}

\subsubsection{Resonant Mollow triplet}

This fitting procedure is first applied on the data of Fig.\,\ref{fig:Fig2_g1t_g1omega_b}. The result is plotted in dash-dotted red and is in very good agreement with the experimental data. The fitted ratio $w_0/\sigma_\mathrm{r}$ gives a radial atomic cloud radius of $\sigma_\mathrm{r} = 950\pm 30\,\mu$m, comparable to the radius measured by absorption imaging and when the laser beams are applied. The fitted broadening of the carrier width due to temperature is of the order of $0.3\,\Gamma$, which, according to Eq.\,(\ref{eq:Delta_omega_Doppler}), corresponds to a temperature of 10\,mK. This temperature is much higher than the one measured on the initial atomic cloud by TOF technique. However, the latter measurement has been realized without the laser beams, and the photons exchanged during the application of these beams significantly heat the atomic sample. When we repeat the TOF measurement after the application of the laser beams, we find a temperature between 10 and 40\,mK. Hence, the temperature increases from 100\,$\mu$K to a few tens of\,mK during the intensity correlations measurements, corresponding to the same order of magnitude extracted from the fit and thus validating the hypothesis that temperature is the main contribution to the carrier broadening.

\begin{figure}[h]
	\centering
	\includegraphics[width=8cm]{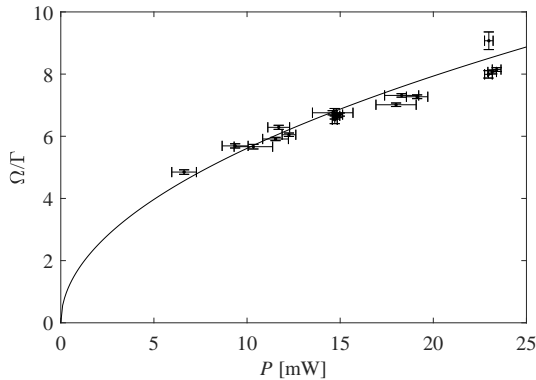}
	\caption{Dots: Rabi frequency extracted from the experimental spectra as a function of the laser power. The vertical error bars are given by the fit of the Mollow triplet spectra and the horizontal error bars come from the power fluctuations during one experimental run. Full curve: Rabi frequency calculated from Eqs.\,(\ref{eq:s}) and (\ref{eq:Omega_Rabi}), and from the measured laser power and waist.}
	\label{fig:Fig3_Omega_P}
\end{figure}

We repeated the same kind of measurements, still at resonance, but with different powers. For each spectrum, we checked that the temperature and $w_0/\sigma_\mathrm{r}$ correspond to what is experimentally extracted by absorption imaging. We finally plot the Rabi frequencies measured in the spectrum as a function of the powers measured just before the vacuum cell (see Fig.\,\ref{fig:Fig3_Omega_P}). The vertical error bars are given by the fit, and the horizontal ones by the power fluctuations estimated during each experimental run. The theoretical Rabi frequency can be deduced from Eqs.\,(\ref{eq:s}) and (\ref{eq:Omega_Rabi}), as well as the laser waist and peak power. This theoretical value corresponds to the curve in Fig.\,\ref{fig:Fig3_Omega_P}, and is in very good agreement with our measurements, the only free parameters being those of the fitting procedure.

\subsubsection{Off-resonance spectrum}

\begin{figure}[h]
	\centering
	\includegraphics[width=8cm]{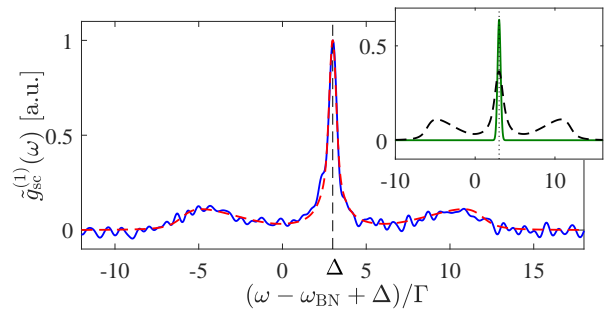}
	\caption{Off-resonant Mollow spectrum of the light scattered by the cold atomic cloud with $\Delta = 3\Gamma$ and $s_0 = 140$. Blue curve: experimental spectrum. The fitting is shown in dashed red, taking into account the elastic and the inelastic contributions. Inset: Green curve: elastic part obtained from the fit; black dashed curve: inelastic part.}
	\label{fig:Fig4_g1t_g1omega_b}
\end{figure}

We now turn to the analysis of the Mollow spectrum when the laser frequency is shifted from the atomic transition by a detuning $\Delta$ of a few linewidths. A typical spectrum is shown in Fig.\,\ref{fig:Fig4_g1t_g1omega_b}, for $\Delta = 3\,\Gamma$ and with $s_0 = 140$. Two sidebands due to the inelastic scattering are still observed. The sharp central part  now corresponds to the inelastic contribution plus a significant elastic component. Because of the lower saturation parameter for a nonzero detuning, the elastic part can no longer be neglected in our analysis. We thus include Eq.~(\ref{eq:el_spectrum}) in the computation of the total spectrum given by \eqref{eq:GaussSpec} where, as before, the finite size of the laser waist and the temperature are accounted for. In particular, the temperature not only broadens the carrier of the inelastic spectrum, but also turns the homogeneous elastic component into a Doppler-broadened Gaussian peak. The resulting fit is plotted in Fig.\,\ref{fig:Fig4_g1t_g1omega_b} in dashed red, showing a good agreement with the experimental data. The temperature and the ratio $w_0/\sigma_\mathrm{r}$ are comparable to the values obtained at resonance, as expected from the large saturation parameters that are used.

\begin{figure}[h]
	\centering
	\includegraphics[width=8cm]{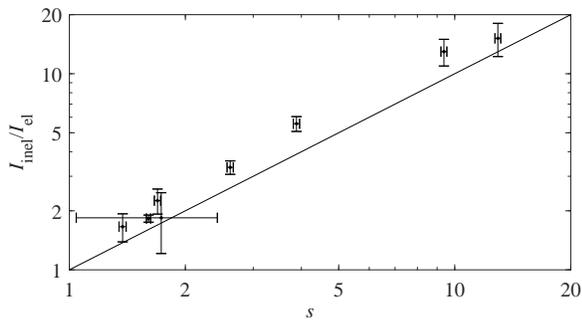}
	\caption{Intensity ratio of the inelastic and elastic part as a function of the saturation parameter. These two quantities are extracted independently from the fit. The intensity ratio is calculated from the integral of the fitted elastic and inelastic spectra, the fitting being done using two different free parameters for the amplitudes of the elastic and inelastic components. The saturation parameter is calculated using the Rabi frequency, also extracted from the fit, and the detuning. Dots: experimental data. Error bars come from the fit. Line: expected behavior given by Eq.\,(\ref{eq:I_inel_I_el}).}
	\label{fig:Fig6_P_inel_P_elas}
\end{figure}

The amplitudes of the elastic component $A_\mathrm{el}$ and of the inelastic component $A_\mathrm{inel}$ are fitted independently as follows:
\begin{equation}
\tilde{g}_\mathrm{sc}{}^{(1)}(\omega) = A_\mathrm{el} \times \tilde{g}_\mathrm{el}{}^{(1)}(\omega) + A_\mathrm{inel} \times \tilde{g}_\mathrm{inel}{}^{(1)}(\omega). \label{eq:total_spectrum_ampl}
\end{equation}
The resulting ratio $I_\mathrm{inel}/I_\mathrm{el}$, corresponding to the ratio of the integral of the inelastic spectrum and the integral of the elastic one, is computed. It is theoretically equal to the saturation parameter, as shown in Eq.\,(\ref{eq:I_inel_I_el}). This measurement has been done for different laser detunings and two different powers. On the other hand, the saturation parameter is calculated using the Rabi frequency $\Omega$ of the inelastic component, also extracted from the fit, thanks to Eq.~\eqref{eq:Omega_Rabi}. As shown in Fig.\,\ref{fig:Fig6_P_inel_P_elas}, the ratio $I_\mathrm{inel}/I_\mathrm{el}$ is in good agreement with the computed saturation parameter.

\subsubsection{Simultaneous pulses \& Non-circular polarization}

\begin{figure}[h]
	\centering
	\includegraphics[width=8cm]{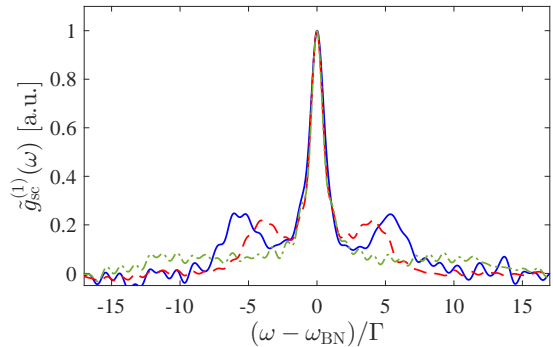}
	\caption{Spectrum of the scattered light when the atomic cloud is illuminated by laser beams A and B one at a time, or both together, in a plane wave or standing wave configuration. Blue continuous curve: laser beam A with circular polarization. Red dashed curve: laser beam A with linear polarization. Green dash-dotted curve: laser beams A and B shined simultaneously and with the same circular polarization.}
	\label{fig:Fig6}
\end{figure}

So far, we have paid attention to have the system as close as possible to an ideal two-level system, with a well defined Rabi frequency. To illustrate the importance of these points, we have performed two more tests. We first measured the spectrum of light scattered by the cold atomic cloud when it is illuminated by a laser beam with a linear polarization, at resonance and with $s_0 \simeq 80$. The result is plotted in Fig.\,\ref{fig:Fig6} (red dashed curve) where we compare the circular and linear polarization. As said before, for a left-handed or right-handed circular polarization, we only excite the transition $F = 3\ (m_F = \pm 3) \rightarrow F'=4\ (m'_F = \pm 4)$. On the contrary, for a linear polarization, different transitions between different Zeeman substates contribute, that possess different dipole moments. The opening of the scattering process to different substates thus results not only in a decrease of the effective Rabi frequency, due to the lower values of the Clebsch-Gordon coefficients compared to the one associated to the $F = 3\ (m_F = \pm 3) \rightarrow F'=4\ (m'_F = \pm 4)$ transition, but also in a broadening of the sidebands and a decrease of their height. In this case, a comparison with the theoretical Mollow triplet requires to compute the steady state populations, their coupling to the excited states with different Clebsch-Gordon coefficient and the scattering rates between different Zeeman sublevels. This last feature is not taken into account by a multiple two-level scheme and thus requires a more evolved modeling, which is out of the scope of this paper.

In order to test the importance of an homogeneous longitudinal Rabi frequency, we illuminated the atomic cloud with a circular polarization, but with both beams with identical circular polarization, shined at the same time. Since the polarization is the same for the two beams, their interference generates a standing wave, that leads to a strong spatial modulation of the Rabi frequency along the beam propagation. Then, as shown in Fig.\,\ref{fig:Fig6} (green dash-dotted curve), the presence of the Mollow sidebands manifests only in a pedestal around the carrier, differently from the sequence with successive pulses. The strong Rabi frequency modulation leads to the superposition of a large range of sidebands, and thus to a blurred signal. Hence, this effect is an extreme case of the broadening of the sidebands due to the beam finite waist, which is responsible for a weaker modulation of the driving amplitude.

\ \\


\section{Conclusion} \label{Sec:Conclusion}

In summary, we have performed the first clear experimental observation of
the resonant Mollow triplet on cold atoms. On resonance, our fluorescence spectrum is completely dominated by inelastically scattered light. The difference between the measurement and theory is fully understood taking into account the finite temperature of our cloud and the finite size of the exciting laser beam. As far as we know, this is the first time that all these effects are integrated in the analysis. Off resonance, the two Mollow sidebands were observed as well, yet in this case the central peak mainly corresponds to Doppler-broadened elastic scattering. The measured ratio of inelastically to elastically scattered light, a quantity that is surprisingly rarely reported in the literature, is in very
good agreement with the theoretical one.

Throughout this work, our cloud was assumed to behave as a large set
of independent emitters, thanks to the low ratio between the optical
thickness (between 1 and 5 on this experiment) and the saturation parameter. An optically denser cloud, for example
achieved by increasing the atom number, could actually allow us to
investigate the effects of the dipole-dipole coupling on the cloud
fluorescence, and in particular the emergence of new resonances. While
few-atom physics requires subwavelength distances to correlate
efficiently the dipole
fluctuations~\cite{Senitzky_1978,Agarwal_1980,BenAryeh_1988}, large
dilute atomic clouds with a large optical thickness have recently been predicted to present an
asymmetric Mollow triplet~\cite{Ott_2013} as well as higher-order
Mollow sidebands~\cite{Pucci_2017}.

Furthermore, the fluorescence of single emitters is known to present
photon correlations~\cite{Aspect_1980,Cohen_1979,Reynaud_1983,Dalibard_1983},
a feature that was used to produce heralded
photons~\cite{Thompson_2006,Ulhaq_2012}. In this context, a promising
idea is to use the dipole-dipole interaction to manipulate the cloud
optical coherence, and thus generate higher-order photon correlations.

\section*{Aknowledgements}

A. E. was supported by a grant from the DGA. R. B. and R. C. T. benefited from Grants from São Paulo Research Foundation
(FAPESP) (Grant Nos. 2015/50422-4  and 2014/01491-0). L. O. G., D. F. S., R. B., M. F., R. K. and R. C. T. received support from project CAPES-COFECUB (Ph879-17/CAPES 88887.130197/2017-01). This experiment was supported by grants from the Excellence Initiative UCA-JEDI from University C\^ote d'Azur. Part of this work was performed in the framework of the European Training network ColOpt, which is funded by the European Union Horizon
2020 program under the Marie Skodowska-Curie action, grant agreement
72146. The Titan X Pascal used for this research was donated by the NVIDIA Corporation.

\bibliography{Biblio} 

\end{document}